\documentclass[showpacs,twocolumn,prb,eqsecnum]{revtex4}

\usepackage{amsmath}
\usepackage{epsfig,color}
\usepackage{graphicx}
\usepackage{dcolumn}
\usepackage{bm}
\newcommand{\nn}{\nonumber}
\newcommand{\beq}{\begin{equation}}
\newcommand{\eeq}{\end{equation}}
\newcommand{\bea}{\begin{eqnarray}}
\newcommand{\eea}{\end{eqnarray}}
\newcommand{\bwt}{\begin{widetext}}
\newcommand{\ewt}{\end{widetext}}

\newcommand{\kf}{k_{\mathrm{F}}}

\newcommand{\bk}{{\bf k}}
\newcommand{\bp}{{\bf p}}
\newcommand{\bq}{{\bf q}}

\begin{document}
\title{Electron self-energy 
near a nematic  quantum critical point}
\author{Markus Garst$^{1}$ and Andrey V. Chubukov$^{2}$}
\affiliation{
$^{1}$Institut fur Theoretische Physik, Universitat zu K\"oln, 50937, K\"oln, Germany \\
and Physik Department, Technische Universit\"at M\"unchen, 85748 Garching, Germany\\
$^{2}$Department of Physics, University of Wisconsin-Madison, 1150 University
Ave., Madison, WI 53706-1390}
\date{\today}
\begin{abstract}
We consider an isotropic Fermi liquid in two dimensions
 near the $n=2$ Pomeranchuk instability in the charge channel.
The order parameter is a quadrupolar stress tensor with two bosonic shear modes with polarizations longitudinal and transverse to the quadrupolar momentum tensor. Longitudinal and transverse bosonic modes are characterized by dynamical exponents $z_\parallel=3$ and $z_\perp=2$, respectively.  Previous studies have found that such a system exhibits multiscale quantum criticality with two different energy scales $\omega \sim \xi^{-z_{\parallel,\perp}}$, where $\xi$ is the correlation length. 
We study the impact of the multiple energy scales on the electron Green function. The interaction with the critical $z_\parallel =3$ mode is known to give rise to a local self-energy that develops a non-Fermi liquid form, $\Sigma(\omega) \sim \omega^{2/3}$ for frequencies larger than the energy scale $\omega \sim \xi^{-3}$.
We find that the exchange of transverse $z_\perp=2$ fluctuations leads to a logarithmically singular renormalizations of the quasiparticle residue $Z$ and the vertex $\Gamma$. We derive and solve renormalization group equations for the flow of $Z$ and $\Gamma$ and show that the system develops an anomalous dimension at the nematic quantum-critical point (QCP). 
As a result, the spectral function at a fixed $\omega$ and varying $k$ has a non-Lorentzian form.  Away from the QCP, we find that the flow of $Z$ is cut at  the energy  scale $\omega_{\rm FL} \propto \xi^{-1}$, associated with the $z=1$ dynamics of electrons.
The $z_\perp=2$ energy scale, $\omega \sim \xi^{-2}$, affects the flow of $Z$ only  if one includes into the theory self-interaction of transverse fluctuations.
\end{abstract}
\maketitle

\section{Introduction}
\label{intro}

The behavior of Fermi liquids (FL) near Pomeranchuk instabilities
attracted high interest in the last few years because of a generic
theoretical interest and potential applications for the cuprates\cite{hinkov}
and ruthenates.\cite{sr} In the FL notations, a Pomeranchuk
instability occurs when one of the harmonics $g_{a,n}$ of the Landau
quasiparticle interaction function approaches $-1$ ($a =c,s$ stands
for charge or spin, and $n$ is the value of the angular momentum).
Examples of Pomeranchuk instabilities include phase separation
$(g_{c,0} =-1$), a ferromagnetic transition ($g_{s,0} =-1$), at which
the Fermi surfaces (FS) of spin-up and spin-down fermions split apart, and
nematic-type transitions in the charge
\cite{metzner_halboth,yamase_1,Oganesyan,MRA,Kee,neto,andy,woelfle,senthil,woelfle_2,charge_we,review}
and spin channels, \cite{hirsch,fradkin,spin_we} which lower the
rotational symmetry of the FS.

The subject of this paper is the $d-$wave ($n=2$) charge nematic instability in an isotropic Fermi liquid at $T=0$, i.e., a charge nematic quantum-critical point (QCP). A charge Pomeranchuk instability in the $d-$wave charge channel was
 first introduced by Halboth and Metzner\cite{metzner_halboth} and Yamase and Kohno~\cite{yamase_1} in the context of the RG analysis of potential instabilities of a 2D Hubbard model. The $d-$wave instability in an isotropic system was analyzed by Oganesyan {\it et al.}\cite{Oganesyan} and in a number of later papers.\cite{Kee,andy,woelfle,senthil,neto,woelfle_2,charge_we} The order parameter for a $d-$wave charge nematic transition is the expectation value of the quadrupolar electron density $c^{\dagger}_{\alpha} Q_{ij}c_{\alpha}$, where $Q_{ij} =
\delta_{ij} \nabla^2 - 2 \partial_i \partial_j$, and $i,j =x,y$. It can be interpreted as a traceless quadrupolar stress tensor
representing elastic shear modes of the Fermi surface (FS)
 (Refs.~\onlinecite{Oganesyan,woelfle_2}). In spatial dimensions $d=2$,
there exist two bosonic shear modes with polarizations longitudinal and transverse to the quadrupolar momentum tensor
 ($\cos 2 \phi$ and $\sin 2\phi$ terms). These two modes are characterized by different dynamics.\cite{Oganesyan} The longitudinal mode is Landau-overdamped by particle-hole pairs and has a dynamical exponent $z_\parallel = 3$. The transverse mode, on the other hand, remains undamped with $z_\perp = 2$. Each mode has an associated ``bosonic mass-shell''
 energy scale, $\omega^\parallel_{\rm FL} \sim \xi^{-z_\parallel}$ and $\omega^\perp_{\rm FL} \sim \xi^{-z_\perp}$, related to the correlation length, $\xi$.
The nematic instability is thus a quantum phase transition with multiple energy scales. 

Multiscale criticality of the nematic transition was recently analyzed by Zacharias {\it et al.} (Ref. \onlinecite{woelfle_2}) within a bosonic Ginzburg-Landau $\phi^4$-theory. These authors considered the interplay of the two modes and the resulting manifestations of the multiple energy scales in thermodynamics.
Longitudinal fluctuations have larger dynamical exponent, $z_\parallel = 3$, 
i.e., larger phase space, and dominate the critical specific heat $C(T)$,
which undergoes a crossover from a FL behavior at  $T < \omega^\parallel_{\rm FL}$ to $C(T) \sim T^{2/3}$ for $T > \omega^\parallel_{\rm FL}$.
Transverse fluctuations with $z_\perp = 2$ have smaller phase space and 
account only for subleading corrections to $C(T)$.  
At the same time, the effective dimension of the transverse mode 
$d+z_\perp = 4$ is upper critical, and its self-interaction gives rise to singular logarithmic corrections to thermodynamics.  As a result, critical thermodynamics becomes sensitive to the second energy scale $\omega_{\rm FL}^\perp$ as well. In particular, Zacharias {\it et al.} argued that the temperature dependence of the correlation length $\xi$ changes qualitatively at a temperature $T \sim \omega_{\rm FL}^\perp$.

In the present work we evaluate the electron self-energy, $\Sigma(\bk,\omega_m)$, close to the nematic transition focusing on the interaction with transverse fluctuations. The self-energy correction, $\Sigma_\parallel$, due to the exchange of longitudinal bosons has been calculated earlier.\cite{Oganesyan,MRA,Kee,woelfle,charge_we} Its most singular part only depends on frequency and is thus purely local. For frequencies $|\omega_m| < \omega_{\rm FL}^\parallel$, $\Sigma_\parallel$ is linear in $\omega_m$ as for a FL liquid but obeys $\Sigma_\parallel(\omega_m) \sim \omega_m^{2/3}$ for higher frequencies. The $\omega_m^{2/3}$ behavior of the self-energy is directly related to $T^{2/3}$ behavior of the specific heat.
Recent advanced studies of longitudinal fluctuations at a nematic transition 
 found that (i) there is infinite number of $\omega_m^{2/3}$ terms with relative coefficients $\mathcal{O}(1)$, even if the theory is extended to large $N$ (the planar diagrams, Ref.~\onlinecite{sslee}) such that there is no guarantee that $\omega_m^{2/3}$ behavior survives, and (ii) there are singular logarithmical corrections at third and higher-orders of the loop expansion.\cite{Metlitski10,Senthil10} In this paper we show that another set of logarithmically 
 singular corrections arises already in lowest loop order 
due to interaction with transverse fluctuations. 
We show that the latter renormalizes the residue of the fermionic propagator contributing to the anomalous dimension for the fermions. 

The issue of renormalizations by transverse fluctuations is tricky. As the contribution of the transverse fluctuations to $C(T)$ is negligible except for the renormalization of the correlation length, one might expect that the electron self-energy remains unaffected by the interaction with the transverse mode. Indeed, we find that the self-energy is smooth at the transverse ``mass-shell'' 
energy scale $\omega_{\rm FL}^\perp$.  We argue, however, that the interaction of electrons with 
transverse fluctuations brings about a third energy scale $\omega_{\rm FL} \sim v_F \xi^{-1}$, where $v_F$ is  the Fermi velocity. This scale corresponds to $z=1$ fermionic excitations near a fermionic mass-shell. We find that above this scale the exchange of transverse bosons gives rise to a singular logarithmic correction to the residue $Z$ of the electron Green's function. 
Summing up the leading logarithmic contributions, we obtain that they exponentiate and contribute to an anomalous dimension 
if the distance to the fermionic mass-shell exceeds this third energy scale $\omega_{\rm FL}$. At criticality,  $\omega_{\rm FL} =0$ and at any distance to the fermionic mass-shell, the electron Green function then acquires the form
\begin{align}\label{IN_1}
G(\bk,\omega_m) \sim \frac{1}{\left(i\, {\rm sign}\, \omega_m |\omega_m|^{2/3} \omega_0^{1/3} - \varepsilon_k \right)^{1-\eta
}},
\end{align}
where $\omega_0 \sim E_F/(k_F a)^4$, $E_F$ and $k_F$ is the Fermi energy and momentum, respectively, and $a$ is a quadrupolar scattering length.
The anomalous dimension, $\eta$, at one-loop order is attributed to the transversal fluctuations, $\eta_{\rm 1L} = \eta_\perp$, and we find $\eta_{\perp} = 1/(2 k_F a)$. 

In order to keep our calculations under control we assumed that the interaction is sufficiently long-ranged in real space such that $a$ is much larger than the inverse average distance between particles, i.e. $\kf a\gg 1$, so that $\eta_\perp$ is small.
It was shown by Metlitski and Sachdev\cite{Metlitski10} 
 and Mross {\it et al.}\cite{Senthil10} that the longitudinal fluctuations also contribute to the anomalous dimension, $\eta$, but this contribution
 appears at third loop-order and is beyond the accuracy of the present study.

The singular behavior of $Z$ affects the spectral function $A(\bk,\omega)$, particularly the 
momentum distribution curve (MDC) measured in ARPES experiments at a fixed, small $\omega$ and varying momentum $\bk$. In the presence of a singular fermionic residue $Z$, the momentum tails of the MDC are no longer of Lorentzian, $1/\varepsilon^2_k$ form, but rather behave as
\begin{align}\label{IN_2}
A(\bk,\omega) \sim \frac{|\omega|^{2/3}}{| \varepsilon_k |^{2-\eta
}}\qquad
{\rm at }\quad |\varepsilon_k| \gg |\omega|^{2/3} \omega_0^{1/3}.
\end{align}
This is the experimentally detectable prediction of the theory.

The input for our calculations is the assumption that a nematic critical point does exist, i.e., that there is no pre-emptive instability at some finite correlation length $\xi$. A pre-emptive pairing instability is always a possibility, but the corresponding $T_c$ is generally quite low.\cite{hasl_p} Metlitski and Sachdev pointed out\cite{Metlitski10} another potential pre-emptive instability: a non-singular correction to the $q^2$ momentum dependence of the static bosonic propagator (which is the same in longitudinal and transverse channels) is large in 
 a large $N$ limit and may lead to a spiral-type 
instability already at a finite $\xi$.\cite{spiral} This issue was further 
 discussed by Mross {\it et al.}\cite{Senthil10} who demonstrated that such
a correction is small and under control if the static bosonic propagator has the form $1/q^{1 + \epsilon}$ instead of $1/q^2$, and $\epsilon$ is small (the 
 non-trivial limit considered in Ref.~\onlinecite{Senthil10} is $\epsilon \ll 1, N \gg 1, \epsilon N = \mathcal{O}(1)$).  We assume in this paper that the regime of small $\epsilon$ extends to $\epsilon =1, N=1$, i.e, that there is no pre-emptive spiral-type instability.     
 
The paper is organized as follows. In the next Section we introduce the effective fermion-boson model for the charge nematic quantum-critical point. In Section~\ref{sec:LongFluc} we analyze the renormalizations arising from the exchange of longitudinal fluctuations and particularly focus on how the longitudinal mode influences the dynamics of the transverse fluctuations. In Section ~\ref{sec:TransFluc} we consider the renormalizations due to transverse fluctuations and derive the result for the transverse self-energy that leads to Eq.~(\ref{IN_1}). The paper concludes with a summary and discussion of the results.

\section{Fermion-boson model for the charge nematic quantum-critical point}

Effective fermion-boson models near quantum-critical points have been
discussed in the literature for nematic~\cite{Oganesyan,Kee,woelfle,woelfle_2} and
other~\cite{acs,spin_we} cases, and we simply state the result: the
proper low-energy theory near a charge QCP is obtained by integrating
out high-energy degrees of freedom and is described by an effective
Hamiltonian with a four-fermion interaction mediated by a static
propagator of soft bosons.  This reflects the fact that the transition
itself and the propagator of soft static bosons are produced by fermions with high energies, of order $E_F$. The fermionic self-energy $\Sigma (\bk,
\omega_m)$ is sensitive to the dynamics of bosons, which by virtue of an energy conservation law comes from low-energy fermions and has to be calculated within the low-energy model, 
self-consistently with $\Sigma(\bk,\omega_m)$.

Because there are two soft boson modes near a nematic transition in $d=2$, the
low-energy theory is a trace of a $2\times 2$
matrix.\cite{Oganesyan,woelfle_2} Non-diagonal terms of this matrix
do not play a role near a QCP and we neglect them. The two
diagonal terms describe interactions mediated by longitudinal and
transverse bosons and are given by 
\begin{align}
\label{suu_1}
&H_{\parallel} =
\\\nn&
\sum_{\bk,\bp,\bq}
\chi_{st}(\bq)
d^{\parallel}_{\bk,\bq}d^{\parallel}_{\bp,\bq}
c^{\dagger}_{\bk+\bq/2,\alpha}c^{\dagger}_{\bp-\bq/2,\beta}c_{\bp+\bq/2,\beta}c_{\bk-\bq/2,\alpha},
\\
\label{suu_2}
&H_{\perp}=
\\\nn&
\sum_{\bk,\bp,\bq} \chi_{st} (\bq) d^\perp_{\bk,\bq} d^\perp_{\bp,\bq}
c^{\dagger}_{\bk+\bq/2,\alpha}c^{\dagger}_{\bp-\bq/2,\beta}c_{\bp+\bq/2,\beta}c_{\bk-\bq/2,\alpha},
\end{align}
where 
\beq
d^{\parallel}_{\bk,\bq}=\sqrt{2}\cos(2\phi_{\bk,\bq}),~~
d^\perp_{\bk,\bq} = \sqrt{2} \sin(2 \phi_{\bk,\bq})
\label{suu_3}
\eeq
are \lq\lq$d-$wave\rq\rq formfactors, 
and the angles, $\phi_{\bk,\bq} = \angle(\bk,\bq)$ and $\phi_{\bp,\bq} = \angle(\bp,\bq)$, are between the bosonic momentum $\bq$ and the two fermionic momenta, $\bk$ and $\bp$, respectively. We assume, as in earlier works,\cite{acs,spin_we}
 that the static bosonic propagator $\chi_{st}(\bq)$ 
is analytic at small $q$ and is given by 
\beq
\chi_{st} (\bq) = \frac{\chi_0}{1 + g_{c,2} + (a q)^2 + \cdots} 
\label{suu_4}
\eeq 
where $\chi_0$ is
the product of the square of electron-boson coupling and the 2D electron
density of states, $\nu = m/\pi$, (within RPA, $\chi_0 = 1/\nu$,
Ref.\onlinecite{charge_we}, and for simplicity we assume below that this holds, i.e.,
$\chi_0 \nu =1$).  As we
already mentioned above, we assume that 
the effective quadrupole interaction is sufficiently long-ranged
in real space, so that we can treat $1/(\kf a)$ as a small parameter.

The nematic instability occurs when the Landau parameter $g_{c,2}$ in Eq.~(\ref{suu_4}) reaches $-1$. Using (\ref{suu_4}), we can introduce a correlation length 
\begin{align}
\xi = \frac{a}{\sqrt{1+g_{c,2}}}.
\label{suu_5}
\end{align}
It diverges as the transition is approached, $g_{c,2} \to -1$.

\subsection{Dynamics of critical nematic fluctuations}

\begin{figure}
\includegraphics[width=0.4\textwidth]{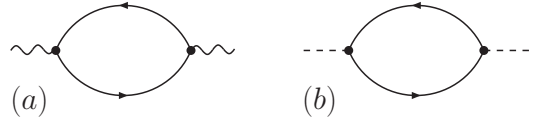}
\caption{Lowest order (a) transverse, $\Pi_\perp$, and (b) longitudinal, $\Pi_\parallel$, polarizations. 
}
\label{fig:Pol0}
\end{figure}

The polarization of electrons gives rise to a retardation of the effective interaction, and, as a result, the susceptibilities in Eqs.~(\ref{suu_1}) and (\ref{suu_2}) acquire  dynamic parts:
\begin{align}
\chi_\alpha (\bq, \Omega_m) = \frac{1}{\nu} \frac{1}{1+g_{c,2} + (a q)^2 + \delta\Pi_\alpha(\bq,\Omega_m)/\nu}
\label{MF_1}
\end{align}
where $\alpha = \parallel,\perp$, and the dynamic part of the polarization bubble is defined as $\delta\Pi_\alpha(\bq,\Omega_m) = \Pi_\alpha(\bq,\Omega_m) - \Pi_\alpha(\bq,0)$. The polarizations for free electrons are given by
\begin{align}\label{MF_2}
\lefteqn{\Pi^{(0)}_{\alpha}(\bq, \Omega_m) =}
\\\nn
& \frac{1}{\beta} \sum_{\bk,\omega_m,\sigma} 
\left(d^\alpha_{\bk,\bq}\right)^2 G_{0,\bk+\bq/2,\omega_m+\Omega_m/2} 
G_{0,\bk - \bq/2,\omega_m - \Omega_m/2},
\end{align}
where $G^{-1}_0(\bk,\omega_m) = i\omega_m - \varepsilon_k$ is
 the bare electron Green function, see Fig.~\ref{fig:Pol0}. 
Interestingly, the dynamics distinguishes between fluctuations that are longitudinal or transverse to the quadrupolar momentum tensor of the electrons.\cite{Oganesyan} To see this, it is convenient to integrate first over $\varepsilon_k$. 
The requirement that the poles of the two Green functions have to be in different half-planes for the $\varepsilon_k$ integration automatically restricts the integration over fermionic frequency, $\omega_m$, to a range of the size of the external bosonic frequency $\Omega_m$. The dynamical $\delta \Pi_\alpha$
 then takes the form
\begin{align}
\delta\Pi^{(0)}_\alpha(\bq, \Omega_m) = \nu \frac{i\Omega_m}{v_F q} \mathcal{F}_\alpha\left(\frac{i \Omega_m}{v_F q}\right),
\label{MF_2_1}
\end{align}
where the functions $\mathcal{F}_\alpha$  are defined as
\begin{align}
\mathcal{F}_{\alpha}(s) = \int_{-\pi}^\pi \frac{d \phi_{\bk,\bq}}{2\pi} 
\frac{\left(d^\alpha_{\bk,\bq}\right)^2}
{s-\cos \phi_{\bk,\bq}}.
\label{MF_3}
\end{align}
Evaluating the angular integrals one obtains
\begin{align}
\mathcal{F}_\parallel(s) &= -2 (2s^2 -1) \left(2 s + \frac{1- 2s^2}{1+s}\sqrt{\frac{s+1}{s-1}} \right),
\label{MF_4_1}
\\
\mathcal{F}_\perp(s) &= -4 s\left(1 - 2 s^2 + 2s(s-1) \sqrt{\frac{s+1}{s-1}}\right).
\label{MF_4_2}
\end{align}
At small $s$, 
\begin{align}
\mathcal{F}_\parallel(s) &= -2 i\, {\rm sign} (Im s) + 4s + ...\\
\mathcal{F}_\perp(s) &= -4s  - 8 i\, {s^2} {\rm sign} (Im s) + ...
\label{MF_4_2_a}
\end{align} 
Both $\mathcal{F}_\parallel(s)$ and $\mathcal{F}_\perp(s)$ contain branch cuts 
 originating from the pole in (\ref{MF_2_1}). In the longitudinal case, the branch-cut
 non-analyticity determines the leading behavior in the limit $|\Omega_m| \ll v_F q$: $\mathcal{F}_\parallel(s) \approx -2 i\ $sign$(Im s)$, giving rise to a dynamical term $|\Omega_m|/(v_F q)$ in $\delta\Pi^{(0)}_\parallel (\bq, \Omega_m)$, characteristic for Landau damping. 
For the transversal polarization, on the other hand, the small frequency limit is analytic:  $\mathcal{F}_\perp(s) \approx -4 s$, because the transverse form factor $d^\perp_{\bk,\bq} = \sqrt{2} \sin (2\phi_{\bk,\bq})$ vanishes when both electrons are on mass-shell, i.e., $\phi_{\bk,\bq} \approx \pi/2$. This limiting behavior can be directly obtained from Eq.~(\ref{MF_3}) by expanding the integrand to first order in $s$, corresponding to the approximation of a quasi-static virtual particle-hole pair. The angular integral in Eq.~(\ref{MF_3}) then averages the direction of the center-of-mass momentum of the pair, $\bk$, over the FS. As a result, the $\Omega_m$ dependence of the transverse polarization $\delta\Pi^{(0)}_\perp (\bq, \Omega_m)$
starts quadratically in frequency, $(\Omega_m/(v_F q))^2$. The branch cut of Eq.~(\ref{MF_4_2}) 
yields a damping term in $\delta\Pi^{(0)}_\perp (\bq, \Omega_m)$ 
only at the third order in $\Omega_m/(v_F q)$. This damping term then can be safely neglected in the critical scaling limit $|\Omega_m| \ll (v_F q)$.

The lowest order polarizations thus yield different dynamics for the transverse and longitudinal 
susceptibilities in the limit $|\Omega_m| \ll v_F q$
\begin{align}
\chi_{0\parallel}(\bq,\Omega_m) &= \frac{1}{\nu} \frac{1}{1+g_{c,2} + (a q)^2 + 2 \frac{|\Omega_m|}{v_F q}},
\label{MF_5_1}
\\
\chi_{0\perp}(\bq,\Omega_m) &= \frac{1}{\nu} \frac{1}{1+g_{c,2} + (a q)^2 + 4 \frac{\Omega_m^2}{(v_F q)^2}}.
\label{MF_5_2}
\end{align}
As a consequence, the theory close to quantum criticality is characterized by multiple dynamical exponents $z$.\cite{Oganesyan} Whereas the non-analytic $\Omega_m$ dependence of the longitudinal mode corresponds to Landau damping characterized by an exponent $z_\parallel=3$, the transversal mode is instead propagating with $z_\perp = 2$. 

\section{Longitudinal fluctuations}
\label{sec:LongFluc}

When electrons scatter off critical nematic fluctuations, singular FL corrections arise.  In the present section, we concentrate on the exchange of longitudinal fluctuations. We first briefly review the results for the ``longitudinal'' self-energy and then discuss how  singular self-energy and vertex corrections due to longitudinal fluctuations affect the polarizations $\delta\Pi_{\alpha}$. 
Of our particular interest here is the renormalization of the transverse $\delta\Pi_{\perp}$ which we then use to calculate the ``transverse'' self-energy, which is the  main subject of our work.

\subsection{Longitudinal self-energy}
\label{sec:LongSigma}

\begin{figure}
\includegraphics[width=0.35\textwidth]{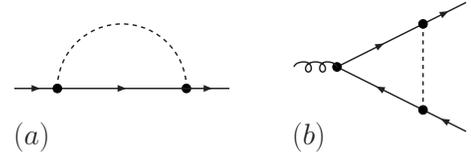}
\caption{Lowest order (a) self-energy and (b) vertex correction due to the exchange of a longitudinal boson (dashed line). The curly line at the vertex represents either a longitudinal or a transverse boson.
}
\label{fig:Sigma&Gamma1}
\end{figure}

The one-loop electronic self-energy arising from the exchange of a longitudinal boson, see Fig.~\ref{fig:Sigma&Gamma1}(a), has been calculated before.\cite{Oganesyan,MRA,charge_we} For completeness, we repeat the result here:
\begin{align}
\Sigma_\parallel(\bk,\Omega_m) = -\frac{1}{\beta} \sum_{\bq, \Omega_m} (d^\parallel_{\bk+\bq/2,\bq})^2 G_{0,\bk+\bq,\omega_m+\Omega_m} \chi_{0\parallel,\bq,\Omega_m}
\label{LF_1}
\end{align}
Neglecting in the form factor $d^\parallel$ 
the bosonic momentum $\bq$ compared to the fermionic $\bk$ and expanding the denominator of $G_0$ in Eq.~(\ref{LF_1}) up to first order in $\bq$, the self-energy takes the form
\begin{align}
\Sigma_{\parallel} (\bk, \omega_m) =&
-\int^{\infty}_0 \frac{dq \,q}{2\pi} \int^{\infty}_{-\infty} \frac{d \Omega_m}{2\pi} \frac{\chi_{0\parallel}(\bq, \Omega_m)}{v_F q} \nonumber\\
&\times\,
\mathcal{F}_\parallel\left(\frac{i(\omega_m + \Omega_m) - \varepsilon_k}{v_F q}\right),
\label{LF_2}
\end{align}
where $\mathcal{F}_\parallel (s)$ is given by (\ref{MF_4_1}). The behavior of $\Sigma_{\parallel} (\bk, \omega_m)$ at low energies is determined by the limiting behavior of the $\mathcal{F}_\parallel$ for small argument: $\mathcal{F}_\parallel(s) \approx -2 i\ $sign$( Im s)$.
Substituting this form, we obtain the local self-energy $\Sigma_\parallel (\omega_m)$,
independent of momentum $\bk$,
\begin{align}
\Sigma_\parallel(\omega_m) = 
\left\{
\begin{array}{ll} 
\lambda i \omega_m & \mbox{if}\,|\omega_m| < \omega^\parallel_{\rm FL} \\
i\, {\rm sign}\, \omega_m |\omega_m|^{2/3} \omega_0^{1/3}& \mbox{if}\, |\omega_m| > \omega^\parallel_{\rm FL},
\end{array}
\right.
\label{LF_3}
\end{align}
where $\omega_0 \sim E_F/(k_F a)^4$. At small frequencies, $\Sigma_\parallel$ has a Fermi-liquid form with the prefactor
\begin{align}
\lambda = \frac{1}{2 (k_F a) \sqrt{1+g_{c,2}}} = \frac{\xi}{2 k_F a^2}
\label{LF_4}
\end{align}
that diverges at criticality. This FL frequency range shrinks, upon approaching criticality, as
\begin{align}
\omega^\parallel_{\rm FL} \sim \frac{E_F}{(k_F a)^4 \lambda^3} \propto \xi^{-3}.
\label{LF_5}
\end{align} 
The dependence $\omega^\parallel_{\rm FL} \propto \xi^{-3}$ is  characteristic for $z_\parallel =3$ dynamics. For frequencies $|\omega_m| > \omega^\parallel_{\rm FL}$, the exchange of a longitudinal boson leads to a non-Fermi liquid, $\omega_m^{2/3}$ form of the self-energy. 

The momentum-dependent part of $\Sigma_\parallel$ comes from higher terms in the expansion of $\mathcal{F}_\parallel (s)$ in $s$ in (\ref{LF_2}) and 
is regular, $\Sigma_\parallel (k, \omega_m =0) \propto \epsilon_k$, 
 and the prefactor is small in $1/(k_Fa)$.~\cite{charge_we}  Three-loop diagrams  give rise to $\epsilon_k \log \epsilon_k$ terms in the self-energy, and eventually have to be taken into consideration.\cite{Metlitski10,Senthil10} 
 For simplicity, we restrict ourselves here to the 
 one-loop longitudinal self-energy Eq.~(\ref{LF_3}). Within this approximation the electron Green's function dressed by longitudinal fluctuations reads
\begin{eqnarray}
G_\parallel(\bk,\omega_m) &=& \frac{1}{i \omega_m + \Sigma_\parallel(\omega_m) - \varepsilon_k} \nonumber \\
&& = \frac{Z_\parallel}{i \omega_m - \bk_F (\bk-\bk_F)/m^*}.
\label{LF_6}
\end{eqnarray}
where by virtue of purely local $\Sigma_\parallel (\omega_m)$, 
$Z_\parallel = m/m^* = [1 + \partial \Sigma_\parallel (\omega_m)/\partial (i \omega_m)]^{-1}$.
In the FL regime, $Z_\parallel = m/m^* = 1/(1 + \lambda)$, in the non-FL regime 
$\omega > \omega^\parallel_{\rm FL}$, both $Z_\parallel$ and $m^*/m$ become functions of frequency.

\subsection{Renormalization of the polarizations}
\label{sec:LongPol}

The lowest-order longitudinal self-energy (\ref{LF_3}) implies the breakdown of the FL close to quantum criticality. The question arises as to how this singular FL correction feeds back into the polarizations and thus modifies the dynamics of the boson propagators, Eqs.~(\ref{MF_5_1}) and (\ref{MF_5_2}). 

Previous studies have established that the Landau damping term of the longitudinal propagator is robust against dressing by longitudinal bosons by two reasons.\cite{charge_we} 
First, dressing fermions in the polarization bubble by longitudinal self-energy adds to the $|\Omega_m|/(v_F q)$ term the overall factor $(Z_\parallel m^*/m)^2$, which remains equal to one even when $Z_\parallel$
 vanishes and $m^*/m$ diverges. Second, vertex corrections need to be evaluated at $k = k_F, \omega_m =\Omega_m =0$, and $q \to 0$, because the Landau damping term is the leading term in an expansion in $\Omega_n/(v_F q)$. In this limit, vertex
 corrections are small in $1/(k_F a)$ 
and are thus irrelevant. As a result, the Landau damping remains unmodified. 

The situation is more tricky for the polarization of the transverse bosons. 
 The dressing of fermions in the transversal polarization $\Pi_{\perp}$ yields for the dynamical $\Omega_m^2/(v_F q)^2$ term an extra factor $Z_\parallel^2 (m^*/m)^3 = 1 + \lambda$ which diverges upon approaching the QCP.\cite{woelfle} At the QCP, the transverse bubble evaluated with dressed fermions has a 
 non-analytic  $|\Omega_m|^{5/3}/(v_F q)^2$ form.
Simultaneously, however, vertex corrections 
 become relevant because in, e.g., the FL regime the leading vertex correction at small but finite 
$\lambda \Omega_m$ is  $\delta\Gamma \sim \Omega_m/(v_F q \cos \phi)$. The 
integrand of $\Pi_\perp$ in Eq.~(\ref{MF_3}) also contains
$\Omega_m \sin^2 {2 \phi}/(v_F q \cos \phi)$ term, and the product of the two yields 
$\Omega_m^2/(v_F q)^2$ term in $\delta \Pi_{\perp}$, which 
 has the same functional form as the term coming from the renormalizations of the electron propagators.
 
This interplay between vertex renormalizations and dressing up of fermions (i.e., self-energy corrections) has been considered perturbatively by Zacharias {\it et al.}
(Ref. \onlinecite{woelfle_2}). They explicitly computed one-loop self-energy
and vertex correction terms with the assumption that the curvature
of the FS can be neglected. They found that these two
corrections exactly cancel each other, i.e., $\delta \Pi_{\perp}$ preserves the same form as for free fermions.
This cancellation is a somewhat unexpected result. Vertex and
self-energy corrections do cancel each other exactly in the opposite
limit $q = 0$ and $\Omega_m \to 0$, as required by the 
Ward identity associated with the conservation of the particle number. However, there is no generic requirement that
there must be a cancellation at arbitrary $\Omega_m/(v_F q)$, and, in
particular, at $|\Omega_m|/v_F q \ll 1$. Yet, a very similar cancellation
between self-energy and vertex corrections to the polarization bubble
at small $\Omega_m/(v_F q)$ has been reported by Kim {\it et al.} for
the problem of fermions coupled to gauge fluctuations.\cite{Kim94}

Below we reconsider this problem with particular emphasis on the role of the curvature of the FS 
and higher-order terms. Although our primary interest is the transverse polarization, for completeness we present calculations for both transverse and longitudinal bubbles.  For the transverse bubble, we show that singular self-energy and vertex corrections cancel, in agreement with Ref.~\onlinecite{woelfle_2}, 
 and the fully renormalized dynamical $\delta \Pi_\perp$ preserves the same  $\Omega^2_m/(v_F q)^2$ form as the bare $\delta \Pi_\perp$. Moreover, the 
the renormalization of the  prefactor is small in $1/(k_F a)$. 
For longitudinal polarization, we indeed find that the leading Landau-damping term is not renormalized. The subleading, $\Omega^2_m/(v_F q)^2$ term is renormalized, but the renormalization is again small in   $1/(k_F a)$. These results  
mean that both longitudinal and transverse fluctuations maintain their FL form and are robust against the dressing of polarizations by longitudinal bosons.
 In the following, we first consider the dressing of $\delta \Pi_{\alpha}$
 by longitudinal bosons in the small $\Omega_m/(v_F q)$ limit. It turns out that in this limit, $\delta \Pi_{\alpha}$ is obtained by a systematic 
 perturbative loop expansion. In a second step, we  evaluate $\delta \Pi_{\alpha}$ in the FL regime
 for an arbitrary value of $\Omega_m/(v_F q)$, by summing up ladder series of vertex correction diagrams.

\subsubsection{Polarization dressed with longitudinal fluctuations, 
small $\Omega_m/(v_F q)$. }

\begin{figure}
\includegraphics[width=0.35\textwidth]{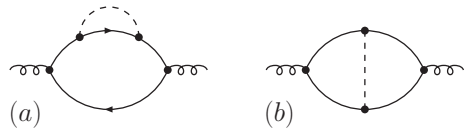}
\caption{Lowest order (a) self-energy and (b) vertex correction to the polarizations from the exchange of a longitudinal boson (dashed line). The outgoing curly lines represent either a longitudinal or transverse boson for the respective $\Pi_\alpha$. Diagram (a) is redundant if the one-loop polarizations in Fig.~\ref{fig:Pol0} are evaluated with dressed Green functions (\ref{LF_6}).}
\label{fig:Pol1}
\end{figure}

We first evaluate 
the polarizations in Fig.~\ref{fig:Pol0} with the renormalized electron Green's function, Eq.~(\ref{LF_6}).  
Substituting (\ref{LF_6}) into (\ref{MF_2}) and integrating over momenta transverse to the FS we obtain, instead of Eq.~(\ref{MF_2_1})
\begin{align} 
&\delta\Pi^{R(\ref{fig:Pol1}a)}_\alpha(\bq, \Omega_m) = 
\frac{i\nu}{v_F q}  \int\limits^{\Omega_m/2}_{-\Omega_m/2} d\omega_m 
\mathcal{F}_\alpha
\Big(
\frac{E_{\Omega_m,\omega_m}}{v_F q} 
\Big)=
\nn\\&
i\nu  \int\limits^{\Omega_m/2}_{-\Omega_m/2} d\omega_m 
\int_{-\pi}^\pi \frac{d\phi_{\bk,\bq}}{2\pi} 
\frac{\left(d^\alpha_{\bk,\bq}\right)^2}{E_{\Omega_m,\omega_m} - v_F q \cos \phi_{\bk,\bq}}, 
\label{LF_7}
\end{align} 
where we introduced the abbreviation $E_{\Omega_m,\omega_m} = i \Omega_m + \Sigma_{\parallel \omega_m+\Omega_m/2} -  \Sigma_{\parallel \omega_m-\Omega_m/2}$,
 and the index $R$ indicates that dressed electron Green functions, $G_\parallel$, are used.

It is convenient to substract from this expression the bare polarization $\delta\Pi^{(0)}_\alpha$ evaluated with the undressed Green's functions. The difference is 
\begin{align} \label{LF_8_1}
&{\delta\Pi^{R(\ref{fig:Pol1}a)}_\alpha(\bq, \Omega_m) - \delta\Pi^{(0)}_\alpha(\bq, \Omega_m) =}
\\\nn&
-\frac{i\nu}{v_F q}  \int\limits^{\Omega_m/2}_{-\Omega_m/2} d\omega_m 
\int_{-\pi}^\pi \frac{d\phi_{\bk,\bq}}{2\pi} \left(d^\alpha_{\bk,\bq}\right)^2
\\\nn&
 \times
\frac{\Sigma_{\parallel \omega_m+\Omega_m/2} -  \Sigma_{\parallel \omega_m-\Omega_m/2}}{(E_{\Omega_m,\omega_m} - v_F q \cos \phi_{\bk,\bq})(i\Omega_m -v_F q \cos \phi_{\bk,\bq})}.
\end{align}
At small $\Omega_m/(v_F q)$ (or, more accurately, at small 
$E_{\Omega_m,\omega_m}/(v_F q)$), 
the frequency-dependent terms in the denominator can be neglected, and evaluating the integral over $\phi_{\bk,\bq}$ we obtain
\begin{align} \label{LF_8}
\lefteqn{\delta\Pi^{R(\ref{fig:Pol1}a)}_\alpha(\bq, \Omega_m) - \delta\Pi^{(0)}_\alpha(\bq, \Omega_m) =}
\\\nn&
b_\alpha \frac{ 4 i \nu}{(v_F q)^2} \int^{\Omega_m/2}_{-\Omega_m/2} d\omega_m
\left[\Sigma_{\parallel \omega_m+\Omega_m/2} -  \Sigma_{\parallel \omega_m-\Omega_m/2}\right],
\end{align}
where $b_\parallel =1$ and $b_\perp = -1$.

This correction can in fact be attributed to the perturbative diagram presented in Fig.~\ref{fig:Pol1}(a). 
 Evaluating the frequency integral, we find that in the FL regime $|\Omega_m| \ll \omega^\parallel_{\rm FL}$, the integral in (\ref{LF_8}) 
is of order $\lambda \Omega^2_m/(v_F q)^2$ and at larger 
 $|\Omega_m| \gg \omega^\parallel_{\rm FL}$ it is of order $|\Omega_m|^{5/3}/(v_F q)^2$.
 Whereas such a non-analytic correction is only subleading for the longitudinal polarization, it would dominate over the $\Omega_m^2/(v_F q)^2$ term of the transversal one in the limit $|\Omega_m| \ll v_F q$. 
However, this singular renormalization coming from the self-energy insertion is canceled out with the renormalization coming from the vertex correction, as we now demonstrate. The polarization bubble with a vertex correction is presented in Fig.~\ref{fig:Pol1}(b). Evaluating this diagram, we obtain 
\begin{align}\label{LF_9}
\Pi^{R(\ref{fig:Pol1}b)}_\alpha(\bq,\Omega_m) =& \frac{1}{\beta}\sum_{\bk,\omega_m,\sigma} 
\left(d^\alpha_{\bk,\bq}\right)^2
\delta \Gamma_{\bk,\omega_m}(\bq,\Omega_m)
\\\nn&
G_{\parallel,\bk+\bq/2,\omega_m+\Omega_m/2} 
G_{\parallel,\bk - \bq/2,\omega_m - \Omega_m/2}.
\end{align}
where $\delta \Gamma_{\bk,\omega_m}(\bq,\Omega_m)$  is shown in Fig.~\ref{fig:Sigma&Gamma1}(b). 
We assume and then verify that $\delta \Gamma$  only depends on the orientation but not on the magnitude of fermionic momentum $\bk$, 
so that we can explicitly integrate over $\varepsilon_k$ in Eq.~(\ref{LF_9}); this automatically restricts the $\omega_m$ integral to a range of size $\Omega_m$. We obtain
\begin{align}\label{LF_10}
&\delta \Pi^{R(\ref{fig:Pol1}b)}_\alpha(\bq,\Omega_m) =
\\\nn& 
i  \nu \int^{\Omega_m/2}_{-\Omega_m/2} d\omega_m 
\int_{-\pi}^\pi \frac{d\phi_{\bk,\bq}}{2\pi} 
\frac{\left(d^\alpha_{\bk,\bq}\right)^2 
\delta \Gamma_{\hat{\bk},\omega_m}(\bq,\Omega_m)}
{E_{\Omega_m,\omega_m} - v_F q \cos \phi_{\bk,\bq}},
\end{align}
We now compute $\delta \Gamma_{\bk,\omega_m}(\bq,\Omega_m)$.  
Neglecting small bosonic compared to large fermionic momenta in form factors
 and using Green functions dressed with the longitudinal self-energy, 
 we obtain 
\begin{align}\label{LF_11}
\lefteqn{\delta \Gamma^{R(\ref{fig:Sigma&Gamma1}b)}_{\bk,\omega_m}(\bq,\Omega_m) =  
\frac{1}{\beta} \sum_{\bq',\Omega'_m} 
\left(d^\parallel_{\bk,\bq'}\right)^2
\chi_{0\parallel}(\bq',\Omega_m')}
\\\nn&\times
G_{\parallel,\bk+\bq'+\bq/2,\omega_m+\Omega_m'+\Omega_m/2}
G_{\parallel,\bk+\bq'-\bq/2,\omega_m+\Omega_m'-\Omega_m/2}.
\end{align}
The integration over the orientation of the bosonic momentum $\bq'$ is dominated by angles for which 
one of the two Green functions is on mass-shell. Keeping only the contributions attributed to these poles Eq.~(\ref{LF_11}) becomes
\begin{gather}
\delta \Gamma^{R(\ref{fig:Sigma&Gamma1}b)}_{\bk,\omega_m}(\bq,\Omega_m) = \frac{4 i}{v_F}
\int\limits^{-\omega_m+\Omega_m/2}_{-\omega_m- \Omega_m/2} \frac{d \Omega'_m}{2\pi} 
\int_0^\infty \frac{dq'}{2\pi} \chi_{0\parallel}(q',\Omega_m') 
\nn\\ \times
\frac{E_{\Omega_m,\omega_m}  - v_F q \cos \phi_{\bk,\bq}}{\left(E_{\Omega_m,\omega_m}- v_F q \cos \phi_{\bk,\bq}\right)^2 - \left(\frac{q' q}{m} \sin \phi_{\bk,\bq}\right)^2}, 
\label{LF_12}
\end{gather}
 The term in the denominator involving $\sin \phi_{\bk,\bq}$ is 
due to the curvature of the FS, i.e., due to the fact that the dispersion $\epsilon_{{\bk_F}+{\bq}} = v_F q \cos \phi_{\bk,\bq} + q^2/(2m)$. 

Expression (\ref{LF_12}) can be further simplified to
\begin{align}\label{LF_13}
&\delta \Gamma^{R(\ref{fig:Sigma&Gamma1}b)}_{\bk,\omega_m}(\bq,\Omega_m) = 
\frac{\Sigma_{\parallel \omega_m+\Omega_m/2} -  \Sigma_{\parallel \omega_m-\Omega_m/2}}{E_{\Omega_m,\omega_m} - v_F q \cos \phi_{\bk,\bq}} 
\\\nn
&
+ \frac{|\Omega_m|}{2 (k_F a)^2} 
\frac{v_F q |\sin \phi_{\bk,\bq}|}{E_{\Omega_m,\omega_m} - v_F q\cos \phi_{\bk,\bq}}
\\\nn&\quad
\times \frac{1}{E_{\Omega_m,\omega_m} - v_F q\cos \phi_{\bk,\bq} + i \frac{v_F q |\sin\phi_{\bk,\bq}|}{2 (k_F a)^2 \lambda} {\rm sign} \Omega_m} 
\end{align}
with $\lambda$ defined in Eq.~(\ref{LF_4}). The first term on the right hand side of Eq.~(\ref{LF_13}) is the result for the vertex correction if we neglect the curvature of the FS. 
It is proportional to the difference of longitudinal self-energies at frequencies of the intermediate fermions. Subsituting this term into (\ref{LF_10}),  neglecting $E_{\Omega_m,\omega_m}$ compared to $v_F q \cos \phi_{\bk,\bq}$ and integrating over $\phi_{\bk,\bq}$, we  find that 
 it exactly cancels out the renormalization of the polarization bubble coming from self-energy insertions, Eq.~(\ref{LF_8}).   

The remaining term in Eq.~(\ref{LF_13}) is attributed to the FS curvature. It is determined by the pole arising from the presence of a finite curvature term in Eq.~(\ref{LF_12}), i.e., by putting the intermediate particle-hole pair on-shell. For its evaluation, one could actually  neglect the frequency dependence of the longitudinal susceptibility, $\chi_{0\parallel}(\bq,\Omega_m)$. The overall $\Omega_m$ dependence is just 
the phase-space to create a particle-hole pair. Substituting this term into (\ref{LF_10}), again negecting $E_{\Omega_m,\omega_m}$ compared to $v_F q \cos \phi_{\bk,\bq}$, and integrating over $\phi_{\bk,\bq}$, we obtain the final result for the leading vertex correction in the limit $|\Omega_m| \ll v_F q$,
\begin{align}\label{LF_15}
\delta \Pi^{R(\ref{fig:Pol1}a\&b)}_\alpha(\bq,\Omega_m) = -b_\alpha \frac{4 \nu}{(k_F a)^2} \frac{\Omega_m^2}{(v_F q)^2},
\end{align}
 where, we remind,  $b_\alpha = 1$ for the longitudinal polarization, 
and $b_\alpha = -1$ for the transverse polarization.
 
Note that to obtain this result one could just neglect the small imaginary part in the denominator in the last line of Eq.~(\ref{LF_13}). The integral over $\phi_{\bk, \bq}$ 
 is then determined by the third-order pole $1/(i 0$ sign$ \Omega_m - v_F q \cos \phi_{\bk,\bq})^3$ in Eq.~(\ref{LF_10}) once we substitute $\delta \Gamma^{R(\ref{fig:Sigma&Gamma1}b)}$ into this formula. This physically corresponds to putting all four intermediate fermions on mass-shell, 
the associated phase-space constraint explains the $(\Omega_m/(v_F q))^2$ dependence in Eq.~(\ref{LF_15}). 

One can straightforwardly check that higher-order vertex correction terms 
only contribute higher-powers of $\Omega_m/v_F q$, i.e., Eq.~(\ref{LF_15}) 
is the full result for the interaction-induced renormalization of the 
polarization bubble  to order $(\Omega_m/v_F q)^2$.
For the longitudinal polarization, this renormalization 
is subleading compared to the $|\Omega_m|/(v_F q)$ Landau-damping term. For the transverse polarization, Eq.~(\ref{LF_15}) has the same  $\Omega_m^2/(v_F q)^2$ dependence as the 
 free-fernmion bubble, Eq.~(\ref{MF_5_2}), i.e., the exchange by longitudinal nematic fluctuations does affect the prefactor for the transverse polarization bubble. Still, the correction to the prefactor is small in $1/(k_F a)^2$, and can therefore be safely neglected.

Note that Eq.~(\ref{LF_15}) is independent of the actual  frequency dependence of the dressed fermionic Green's function, the only thing that matters is that the sign of the dynamic part of the dressed $G^{-1} (\bk, \omega_m)$ is the same as the sign of $\omega_m$. In fact, Eq. (\ref{LF_15}) could be also derived using a quasi-static approximation for the fermion propagator,
 $G^{-1}(\bk,\omega_m) \approx i 0 \ $sign$\omega_m- \varepsilon_k$ or 
 using bare Green functions, $G_0$, instead of $G$.

\subsubsection{Polarizations at arbitrary $\Omega_m/(v_F q)$: ladder approximation}

\begin{figure}
\includegraphics[width=0.35\textwidth]{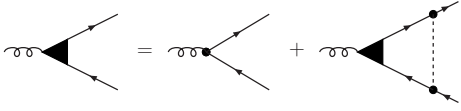}
\caption{Ladder approximation for the vertex summing up the repeated exchange of longitudinal bosons (dashed line). 
The curly line represents either a transverse or a longitudinal boson.}
\label{fig:vertexladder}
\end{figure}

We now extend the analysis to arbitrary $\Omega_m/(v_F q)$. To avoid cumbersome expressions, we restrict the consideration to the FL regime,  $|\Omega_m| \ll \omega_{\rm FL}^\parallel$. The extention to non-FL regime at larger $\Omega_m$ is straightforward~\cite{chub_ward} but requires more efforts. 
 
In the FL regime, we can use the FL form of the longitudinal self-energy, $\Sigma_\parallel(\omega_m) = \lambda i \omega_m$, and the one-loop vertex correction becomes 
\begin{align}\label{LF_16}
\lefteqn{\delta \Gamma^{R(\ref{fig:Sigma&Gamma1}b)}_{\bk,\omega_m}(\bq,\Omega_m) = }
\\\nn&
\frac{i \Omega_m \lambda}{(1+\lambda) i\Omega_m - v_F q \cos \phi_{\bk,\bq} + 
i \frac{ v_F q |\sin \phi_{\bk,\bq}|}{2 (k_F a)^2 \lambda}{\rm sign}\Omega_m}.
\end{align}
Higer-order vertex corrections form series in $(\delta \Gamma^{R(\ref{fig:Sigma&Gamma1}b)}_{\bk,\omega_m}(\bq,\Omega_m))^n$ and remain $\mathcal{O}(1)$ when $\lambda |\Omega_m| > v_F q$. 
We verified that at  $k_F a \gg 1$, which we assume to hold, leading
vertex corrections form a ladder series, while non-ladder terms are
small in $1/(k_F a)$ (Ref.\onlinecite{charge_we}). The ladder diagrams form a geometrical series, see Fig.~\ref{fig:vertexladder}, and can be easily summed up. The resulting vertex in the ladder approximation reads
\begin{align}\label{LF_17}
\lefteqn{\Gamma^{\rm ladder}(\hat{\bk},{\bf q}, \Omega_m) = }
\\\nn&
\lefteqn{ 1 + 
\delta \Gamma^{R(\ref{fig:Sigma&Gamma1}b)}_{\bk,\omega_m}(\bq,\Omega_m)
 + (\delta \Gamma^{R(\ref{fig:Sigma&Gamma1}b)}_{\bk,\omega_m}(\bq,\Omega_m))^2 + ... =}
\\\nn&
\frac{i\Omega_m (1 + \lambda) - v_F q \left( \cos \phi_{\bk,\bq} - i \frac{{\rm sign}\Omega_m}{2 (k_F a)^2 \lambda} |\sin \phi_{\bk,\bq}| \right)
}
{i\Omega_m - v_F q \left( \cos \phi_{\bk,\bq} - i \frac{{\rm sign}\Omega_m}{2 (k_F a)^2 \lambda} |\sin \phi_{\bk,\bq}| \right)
}.
\end{align}
\begin{figure}
\includegraphics[width=0.2\textwidth]{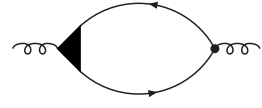}
\caption{Polarization containing a vertex in the ladder approximation, see Fig.~\ref{fig:vertexladder}.
}
\label{fig:LadderPi}
\end{figure}

The polarizations are now obtained from the diagram with dressed fermion lines and the full vertex, see Fig.~\ref{fig:LadderPi},
\begin{align}\label{LF_18}
\Pi_\alpha(\bq, \Omega_m) =
\frac{1}{\beta} &\sum_{\bk,\omega_m,\sigma} 
\left(d^\alpha_{\bk,\bq}\right)^2 \Gamma^{\rm ladder}(\hat{\bk},{\bf q}, \Omega_m)
\\\nn&
G_{\parallel \bk+\bq/2,\omega_m+\Omega_m/2} 
G_{\parallel \bk - \bq/2,\omega_m - \Omega_m/2}.
\end{align}
For its dynamical part we obtain
\begin{align}\label{LF_19}
\lefteqn{\delta \Pi_{\alpha} (\bq, \Omega_m) = i \Omega_m \nu}
\\\nn
&\int_0^{2\pi} \frac{d \phi_{\bk,\bq}}{2\pi} \frac{\left(d^\alpha_{\bk,\bq}\right)^2}{i\Omega_m - 
v_F q \left(\cos \phi_{\bk,\bq} - i \frac{{\rm sign}\Omega_m}{2 (k_F a)^2 \lambda} |\sin \phi_{\bk,\bq}| \right)} \times
\\  \nn
&\left[\frac{i\Omega_m (1 + \lambda) - v_F q \left(\cos \phi_{\bk,\bq} - i \frac{{\rm sign}\Omega_m}{2 (k_F a)^2 \lambda} |\sin \phi_{\bk,\bq}|\right)}{i\Omega_m (1 + \lambda) - v_F q \cos \phi_{\bk,\bq}}\right].
\end{align} 
For $q =0$ and  finite $\Omega_m$, this reduces to the density of states,
 $\delta \Pi_{\alpha} =\nu$, as it should be by the Ward identity. At arbitrary $\Omega_m/(v_F q)$, the full $\delta \Pi_{\alpha}$ can be expressed as the sum of two terms,
\bea\label{LF_20}
\lefteqn{\delta \Pi_{\alpha} (\bq, \Omega_m) =}\\\nonumber&&
 \delta \Pi^{(0)}_{\alpha} (\bq,
\Omega_m) + \frac{4\nu}{(k_F a)^2} \left(\frac{\Omega_m}{v_F q}\right)^2
\mathcal{P}_{\alpha}\left(\frac{i\Omega_m}{v_F q},\lambda\right).
\eea 
Close to the QCP $\lambda$ is large, and 
the functions $\mathcal{P}$ are given by 
\begin{align}\label{LF_21_1}
\mathcal{P}_\alpha(s,\lambda) = \frac{i\, {\rm sign Im} s}{16\pi} \int\limits_{-\pi}^{\pi} d\phi 
\frac{d^2_\alpha(\phi)\ |\sin\phi|}{(s-\cos\phi)^2 (s(1+\lambda) - \cos\phi)},
\end{align}
where, we remind, $s = i \Omega_m/v_F q$, $d_\parallel(\phi) = \sqrt{2} \cos\phi$ and $d_\perp(\phi) = \sqrt{2} \sin\phi$. At small $s$, i.e., at $|\Omega_m| \ll v_F q$, $\lim_{s \to 0} \mathcal{P}_\parallel(s,\lambda) = -1$ and $\lim_{s \to 0} \mathcal{P}_\perp(s,\lambda) = 1$, and  Eq.~(\ref{LF_20}) reproduces the perturbative result (\ref{LF_15}).
At finite $s$, there is an intermediate regime at large $\lambda$ when  
$|s| \ll 1, |s| (1+\lambda) \sim 1$, where 
$\mathcal{P}_\alpha(s,\lambda) \approx \mathcal{P}_\alpha(s \lambda)$ 
becomes a function of $s \lambda$, e.g., $\mathcal{P}_\perp(s \lambda) \propto 1/\sqrt{1 +( i s \lambda)^2}$. At $|s| \ll 1 \ll |s| \lambda$, both $\Pi_\perp$ and $\Pi_\parallel$ behave as $\mathcal{P}_\alpha(s,\lambda) \propto 1/(s\lambda)$, e.g.,  the second term in (\ref{LF_21_1}) has the same $|\Omega_m|/v_F q$, form  as Landau damping, but is suppressed by $1/\lambda$. Finally, for large $s$ we get $\mathcal{P}_\alpha(s,\lambda) \propto 1/(s^3 \lambda)$.

In the non-FL quantum-critical regime the 
computation of $\delta \Pi_{\alpha} (\bq, \Omega_m)$ becomes more complex, but
a comparison with the FL result shows that at $\Sigma_{\parallel} (\Omega_m) \sim v_F q$ and $\Omega_m \ll v_F q$, 
\bea\label{LF_20_a}
\lefteqn{\delta \Pi_{\alpha} (\bq, \Omega_m) =}\\\nonumber&&
 \delta \Pi^{(0)}_{\alpha} (\bq,
\Omega_m) + \frac{4\nu}{(k_F a)^2} \left(\frac{\Omega_m}{v_F q}\right)^2
\mathcal{P}_{\alpha}\left(\frac{\Sigma_{\parallel} (\Omega_m)}{v_F q}\right).
\eea 
The overall conclusion of the analysis in this section is that, as long as we approximate the
self-energy by $\Sigma_{\parallel} (\omega_m)$, 
 the original forms of both polarizations, longitudinal and transverse, are preserved. 
In particular, neglecting quantitative corrections that are small in the parameter $1/(k_F a)$, the full $\chi_{\alpha} (\bq, \Omega_m)$ can be safely approximated by Eqs.~(\ref{MF_5_1}) and (\ref{MF_5_2}). 

\section{Transverse fluctuations}
\label{sec:TransFluc}

\begin{figure}
\includegraphics[width=0.35\textwidth]{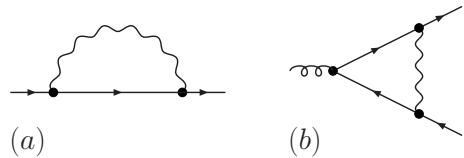}
\caption{Lowest order (a) self-energy and (b) vertex correction due to the exchange of a transverse boson (wiggly line). The curly line at the vertex represents either a longitudinal or a transverse boson.
}
\label{fig:Sigma&Gamma2}
\end{figure}

We now turn to the renormalizations arising from the exchange of critical transverse fluctuations, i.e., bosonic nematic modes with momentum $\bq$ and a polarization transverse to the corresponding quadrupolar momentum tensor, $\delta_{ij} q^2 - 2 q_i q_j$. We first analyze  perturbative one-loop corrections to the electron self-energy and the vertex and demonstrate that they are logarithmically large. Afterwards, we sum up the leading logarithmic singularities and show that the full result is
 the appearance of an anomalous dimension of the electron Green function.

\subsection{One-loop self-energy}

The lowest order electron self-energy arising from the exchange of a transverse boson is shown in Fig.~\ref{fig:Sigma&Gamma2}(a). Following the same steps leading to Eq.~(\ref{LF_2}), we obtain for the transverse self-energy
\begin{align}
&\Sigma^{R(\ref{fig:Sigma&Gamma2}a)}_{\perp} (\bk, \omega_m) =
-\int^{\infty}_0 \frac{dq \,q}{2\pi} \int^{\infty}_{-\infty} \frac{d \Omega_m}{2\pi} \frac{\chi_{0\perp}(\bq, \Omega_m)}{v_F q} \nonumber\\
&\times\,
\mathcal{F}_\perp\left(\frac{i(\omega_m + \Omega_m) + \Sigma_{\parallel \omega_m+\Omega_m}- \varepsilon_k}{v_F q}\right),
\label{TF_1}
\end{align}
where  $\mathcal{F}_\perp$  is given by Eq.~(\ref{MF_4_2}), and
the index $R$ again indicates that electron Green functions dressed with $\Sigma_\parallel$ were used for its evaluation. Consider first the contribution obtained by setting external energy and momentum exactly on mass-shell, $i\omega_m + \Sigma_\parallel(\omega_m) - \varepsilon_k = 0$. In the low-energy limit, we approximate $\mathcal{F}_\perp$ by its value at a small argument and obtain
\begin{align}\label{TF_2}
&\left.\Sigma^{R(\ref{fig:Sigma&Gamma2}a)}_{\perp} (\bk, \omega_m)\right|_{\rm mass\ shell} =
\\\nn&
4\int^{\infty}_0 \frac{dq \,q}{2\pi} \int^{\infty}_{-\infty} \frac{d \Omega_m}{2\pi} 
\frac{\chi_{0\perp}(\bq, \Omega_m)}{(v_F q)^2} 
\left(\Sigma_{\parallel \omega_m+\Omega_m} - \Sigma_{\parallel \omega_m}\right).
\end{align}
The integration over $q$ is straighforward and performing it and then integrating over frequency, we obtain that one-loop transverse self-energy on the mass shell behaves as 
 $\omega_m \lambda^{2/3}/(k_F a)^{4/3}$ 
 at the lowest energies and as $\omega_m^{2/3} \omega_0^{1/3}/(k_F a)$ at the QCP.
(In both limits the $\Omega_m$ integral samples the frequency regime where the longitudinal self-energy has the non-FL form, $\Sigma_\parallel(\omega_m) \sim |\omega_m|^{2/3}$.) This additional self-energy correction is smaller than $\Sigma_{\parallel} (\omega_m)$ both in the FL regime and at the QCP, although at the QCP the relative smallness is only in $1/(k_F a)$.  We see therefore that mass-shell transverse self-energy is essentially irrelevant. In fact, we will show in Sec.~\ref{sec:Mixed2Loop} below that the full transverse self-energy on the mass shell is even smaller as $\left.\Sigma^{R(\ref{fig:Sigma&Gamma2}a)}_{\perp} (\bk, \omega_m)\right|_{\rm mass\ shell}$ is actually canceled by other terms. 

The important transversal self-energy (\ref{TF_1}) is its off-shell part. Expanding Eq.~(\ref{TF_1}) in the distance to the mass-shell, $\delta \Sigma = \Sigma - \Sigma|_{\rm mass\, shell}$, we obtain
\begin{align}
&\delta\Sigma^{R(\ref{fig:Sigma&Gamma2}a)}_{\perp} (\bk, \omega_m) =
-\left(i \omega_m + \Sigma_{\parallel}(\omega_m)- \varepsilon_k\right)
\int\limits^{\infty}_{q_{\rm min}} \frac{dq \,q}{2\pi}  
\nn\\&\times
\int\limits^{\infty}_{-\infty} \frac{d \Omega_m}{2\pi}
\frac{\chi_{0\perp}(\bq, \Omega_m)}{(v_F q)^2}
\mathcal{F}'_\perp
\Big(
\frac{i\Omega_m + \Sigma_{\parallel \omega_m+\Omega_m}- \Sigma_{\parallel \omega_m}}{v_F q}\Big)
.
\label{TF_3}
\end{align}
where $\mathcal{F}'_\perp$ is a derivative of $\mathcal{F}_\perp$ 
 with respect to its argument.
The frequency integral is determined by the pole of $\chi_{0\perp}$ and the remaining momentum integral is logarithmically large.
 To logarithmic accuracy, the lower boundary of the momentum integral is determined by the distance to the fermionic mass-shell $v_F q_{\min} \simeq |i \omega_m + \Sigma_{\parallel}(\omega_m) - \varepsilon_k|$ such that an expansion of Eq.~(\ref{TF_1}) remains justified. At large momenta, the integral is cut
  by the $\mathcal{F}_\perp$ function. Effectively, the upper boundary is given by $q_{\rm max}\simeq 1/a$ where $\Omega_m/(v_F q)$, taken at the bosonic pole, becomes of order one.
In this momentum and frequency range, $\Omega_m/(v_F q)$ dominates over the difference of longitudinal self-energies divided by $v_F q$ in the argument of the $\mathcal{F}'_\perp$ function, the latter being at most of order $1/(k_F a) \ll 1$. So we can use a small-argument expansion for  $\mathcal{F}'_\perp$ with $\mathcal{F}'_\perp(0) = - 4$.  We then obtain, with logarithmic accuracy
\begin{align}\label{TF_4}
&\delta\Sigma^{R(\ref{fig:Sigma&Gamma2}a)}_{\perp} (\bk, \omega_m) = 
-\frac{i \omega_m + \Sigma_{\parallel}(\omega_m)- \varepsilon_k}{8 k_F a} \mathcal{F}'_\perp(0)
\\\nn&\quad\times
 \int\limits^{1/a}_{q_{\rm min}} \frac{dq}{\sqrt{q^2 + \xi^{-2}}}  =
\frac{i \omega_m + \Sigma_{\parallel}(\omega_m) - \varepsilon_k}{2 k_F a}
\\\nn&\quad\times
\log\left[
\frac{E_F/(k_F a)}{
{\rm max}\left\{
|i \omega_m + \Sigma_{\parallel}(\omega_m) - \varepsilon_k|, \omega_{\rm FL} \right\}}
\right].
\end{align}
The transversal self-energy thus yields a logarithmically singular correction to the fermionic residue $Z$.  The  logarithm is cut-off at the scale,
\begin{align}\label{TF_5}
\omega_{\rm FL} \sim \frac{E_F}{(k_F a)^2 \lambda} \sim  v_F \xi^{-1}.
\end{align}
The relation $\omega_{\rm FL} \propto \xi^{-1}$ is characteristic for the $z=1$ dynamics of electrons. 
The corresponding  momentum scale $q_{\rm FL} = \omega_{\rm FL}/v_F \propto \xi^{-1}$.  
The $z=1$ scaling actually comes from the upper limit of the logarithmical integral for which, as we said, $\Omega_m/(v_F q) \sim (aq)^2 = \mathcal{O}(1)$.
 
\subsection{One-loop vertex correction}

We next show that not only the self-energy but also the vertex correction
 due to the exchange of a transverse boson  is logarithmically enhanced. 
The vertex correction is presented in Fig.~\ref{fig:Sigma&Gamma2}(b). 
It is given by the expression similar to Eq.~(\ref{LF_11}),
 but with transverse form factors and the transverse susceptibility instead 
 of the longitudinal ones. The leading contribution to the transvese vertex 
correction again
comes from the regime where $\Omega_m + \Sigma_{\parallel} (\Omega_m) \ll v_F q$, and
 the two fermionic Green's functions can be approximated by their static forms $G_{\bk+\bq,\omega_m} \approx 1/(-v_F q \cos \phi_{\bk, \bq})$. Substituting these forms into the vertex correction diagram, we obtain
\begin{align}\label{TF_6}
\delta \Gamma^{R(\ref{fig:Sigma&Gamma2}b)}_{\bk,\omega_m}(\bq,\Omega_m) =&  
\int_{-\infty}^{\infty} \frac{d\Omega_m'}{2\pi} \int_{q_{\rm min}}^{q_{\rm max}} 
\frac{d q' q'}{2\pi}\chi_{0\perp}(q',\Omega_m')
\nn\\&
\times
\int_{-\pi}^\pi \frac{d \phi}{2\pi} \frac{2 \sin^2 2\phi}{(v_F q' \cos \phi)^2}.
\end{align}
This momentum regime is bounded from below by the distance to the two fermionic mass shells of the internal fermions,  $v_F q_{\rm min} \simeq {\rm max}\left\{|i \omega_m \pm i\Omega_m/2 + \Sigma_{\parallel \omega_m \pm \Omega_m/2}- \varepsilon_{\bk\pm \bq/2}|\right\}$. The upper bound $q_{\rm max} \simeq 1/a$ again ensures that the internal bosonic energy taken at its pole
remains sufficiently small. The remaining integral in Eq.~(\ref{TF_6}) is logarithmically large,
\begin{align}\label{TF_7}
\lefteqn{\delta \Gamma^{R(\ref{fig:Sigma&Gamma2}b)}_{\bk,\omega_m}(\bq,\Omega_m) =  \frac{1}{2 k_F a} \times}
\\\nn&
\log\left[
\frac{E_F/(k_F a)}{
{\rm max}\left\{
| i \omega_m \pm i \Omega_m/2 + \Sigma_{\parallel \omega_m \pm \Omega_m/2}- \varepsilon_{\bk\pm \bq/2}|,
\omega_{\rm FL} \right\}}
\right]
\end{align}
where $\omega_{\rm FL}$ is defined in Eq.~(\ref{TF_5}).

\begin{figure}
\includegraphics[width=0.4\textwidth]{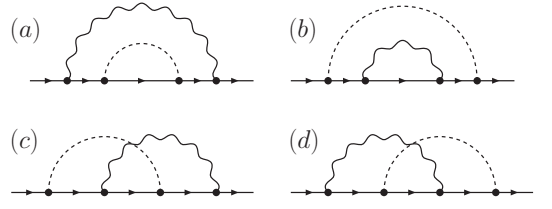}
\caption{Two loop fermionic self-energy diagrams with the exchange of a transverse (wiggly line) and longitudinal boson (dashed line). Diagram (a) is redundant if the one-loop diagram in Fig.~\ref{fig:Sigma&Gamma2}(a) is evaluated with a dressed electron Green function (\ref{LF_6}).
}
\label{fig:2LoopSigma1}
\end{figure}

\subsection{Two-loop corrections}
 
In the next loop-order, there are mixed diagrams that involve transverse and longitudinal bosons, and diagrams with two transverse bosons. For the self-energy, we demonstrate that mixed diagrams just cancel the dominant term in the on-shell one-loop self-energy, Eq.~(\ref{TF_2}). We show that the self-energy and the vertex corrections containing two transverse boson exchanges contribute $\log^2$ terms. We also discuss the cancellation of singular logarithms in the polarizations.

\subsubsection{Two-loop self-energy with transverse and longitudinal boson exchange}
\label{sec:Mixed2Loop}

Consider the two-loop diagrams to the self-energy in Fig.~\ref{fig:2LoopSigma1} obtained by the exchange of a single transverse and a single longitudinal boson. Diagram (a) can be disregarded; it is redundant as we already evaluated the one-loop diagram in Fig.~\ref{fig:Sigma&Gamma2}(a) with a dressed Green function $G_\parallel$. So we need to evaluate the remaining three: diagram (b) is a longitudinal self-energy diagram with a transversal self-energy insertion, and (c) and (d) are self-energy diagrams with vertex corrections. 

Consider first the diagram in Fig.~\ref{fig:2LoopSigma1}(b). We have 
\begin{align}\label{TF_2_1}
\Sigma^{R(\ref{fig:2LoopSigma1}b)}(\bk,\Omega_m) = \frac{1}{\beta} \sum_{\bq, \Omega_m} & (d^\parallel_{\bk,\bq})^2 
\left(G_{\parallel,\bk+\bq,\omega_m+\Omega_m}\right)^2 
\\\nn& \times
\Sigma^{R(\ref{fig:Sigma&Gamma2}a)}_{\perp \bk+\bq,\omega_m+\Omega_m}  
\chi_{0\parallel,\bq,\Omega_m}.
\end{align}
Instead of using the result Eq.~(\ref{TF_4}) for $\Sigma^{R(\ref{fig:Sigma&Gamma2}a)}_\perp$ here, it is actually more convenient to perform first the angular integration over the orientation of bosonic momentum $\bq$. The resulting expression can then be approximated as
\begin{align}\label{TF_2_2}
&\Sigma^{R(\ref{fig:2LoopSigma1}b)}(\bk,\Omega_m) = 
\\\nn& 
4\int^{\infty}_0 \frac{dq \,q}{2\pi} \int^{\infty}_{-\infty} \frac{d \Omega_m}{2\pi} 
\frac{\chi_{0\perp}(\bq, \Omega_m)}{(v_F q)^2} 
\left(\Sigma_{\parallel \omega_m+\Omega_m} - \Sigma_{\parallel \omega_m}\right).
\end{align}
The two diagrams in Fig.~\ref{fig:2LoopSigma1}(c) and (d) are equivalent and can be expressed as 
\begin{align}\label{TF_2_3}
&\Sigma^{R(\ref{fig:2LoopSigma1}c\&d)}(\bk,\Omega_m) = 
-\frac{2}{\beta} \sum_{\bq, \Omega_m} \left( d^\perp_{\bk,\bq} \right)^2
\\\nn&
\delta\Gamma^{R(\ref{fig:Sigma&Gamma1}b)}_{\bk+\bq/2,\omega_m+\Omega_m/2}(\bq,\Omega_m) 
G_{\parallel,\bk+\bq,\omega_m+\Omega_m} \chi_{0\perp,\bq,\Omega_m}.
\end{align}
where $\delta\Gamma^{R(\ref{fig:Sigma&Gamma1}b)}_{\bk+\bq/2,\omega_m+\Omega_m/2}(\bq,\Omega_m)$ is given by (\ref{LF_13}). To leading order in $1/(k_F a)$, we can neglect the curvature and approximate this vertex by the first term on the right hand side of (\ref{LF_13}).
To the same accuracy, we can neglect $E_{\Omega_m,\omega_m}$ in the denominator of (\ref{LF_13}) (i.e., approximate the propagator of an 
intermediate particle-hole pair in the vertex correction by a static limit), 
and also approximate the fermion Green's function by its static limit. 
Applying these approximations, we obtain
\begin{align}\label{TF_2_4}
&\Sigma^{R(\ref{fig:2LoopSigma1}c\&d)}(\bk,\Omega_m) = 
\\\nn&
-8\int^{\infty}_0 \frac{dq \,q}{2\pi} \int^{\infty}_{-\infty} \frac{d \Omega_m}{2\pi} 
\frac{\chi_{0\perp}(\bq, \Omega_m)}{(v_F q)^2} 
\left(\Sigma_{\parallel \omega_m+\Omega_m} - \Sigma_{\parallel \omega_m}\right).
\end{align}
As announced, the sum of the two-loop contributions, Eqs.~(\ref{TF_2_2}) and (\ref{TF_2_4}), cancels  the leading term in the dressed one-loop transverse self-energy  on the mass-shell, Eq.~(\ref{TF_2}). The subleading terms are not canceled, but they are small and of no relevance. 

\subsubsection{Two-loop self-energy with exchange of two transverse bosons}
\label{sec:2LoopSigma}

\begin{figure}
\includegraphics[width=0.4\textwidth]{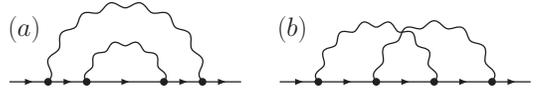}
\caption{Two loop fermionic self-energy diagrams with the exchange of two transverse bosons (wiggly line).
}
\label{fig:2LoopSigma2}
\end{figure}

We now turn to the two-loop diagrams for the self-energy containing two transverse boson exchanges, see Fig.~\ref{fig:2LoopSigma2}. We verified that, to logarithmic accuracy,  the self-energy diagram (a) is dominated by the regime where the momentum of the inner transverse boson is larger than the momentum of the outer one. Using the one-loop result (\ref{TF_4}) for the internal self-energy, we obtain
\begin{align}
&\Sigma^{R(\ref{fig:2LoopSigma2}a)}(\bk,\Omega_m) = \frac{1}{2 k_F a}
\frac{1}{\beta} \sum_{\bq, \Omega_m} (d^\perp_{\bk,\bq})^2 G_{\parallel \bk+\bq,\omega_m+\Omega_m} \chi_{0\perp \bq,\Omega_m}
\nn\\&\times
\log\left[
\frac{E_F/(k_F a)}{
{\rm max}\left\{
|i \omega_m + i \Omega_m + \Sigma_{\parallel \omega_m+\Omega_m} - \varepsilon_{\bk+\bq}|, \omega_{\rm FL} \right\}}
\right].
\label{TF_2_5}
\end{align}
Diagram (b) has two dominant contributions: from the regime where  the first boson is larger than the second one and  vice versa. Both contributions can be 
re-expressed via the one-loop vertex correction due to transverse boson exchange. Using Eq.~(\ref{TF_7}) for this vertex correction we obtain
\begin{align}
&\Sigma^{R(\ref{fig:2LoopSigma2}b)}(\bk,\Omega_m) = 
-\frac{1}{k_F a}\frac{1}{\beta} \sum_{\bq, \Omega_m} (d^\perp_{\bk,\bq})^2 G_{\parallel \bk+\bq,\omega_m+\Omega_m} 
\chi_{0\perp \bq,\Omega_m}
\nn\\&\times
\log\left[
\frac{E_F/(k_F a)}{
{\rm max}\left\{
|i \omega_m + i \Omega_m + \Sigma_{\parallel \omega_m+\Omega_m} - \varepsilon_{\bk+\bq}|, \omega_{\rm FL} \right\}}
\right].
\label{TF_2_6}
\end{align}
Observe that the ``vertex'' diagram (b) thus gives twice the contribution of the ``self-energy'' diagram (a), see Eq.~(\ref{TF_2_5}), and is of opposite sign. The two diagrams then partially cancel each other, and the full result is one half of $\Sigma^{R(\ref{fig:2LoopSigma2}b)}(\bk,\Omega_m)$.

For the remaining integration, we  focus on the off-shell part and expand the Green's function to linear order in the distance to the mass-shell. The integration over frequency $\Omega_m$ is dominated by the bosonic pole of the transverse susceptibility. The leading contribution to the momentum integral comes from the region where the momentum component, $q_\perp$, perpendicular to the external fermionic momentum is much larger than $\xi^{-1}$. 
 In this regime, the integrand depends on the longitudinal component, $q_\parallel$, only via the form factor, $(d^\perp_{\bk,\bq})^2 = 8 q^2_\perp q^2_\parallel/(q^2_\perp + q^2_\parallel)^2$, and the integration over $q_\parallel$ is easily performed. The final integration over $q_\perp$ then reduces to 
$\int d q_\perp/q_\perp \log q_\perp$. Collecing the prefactors, we then obtain a $\log^2$ correction to the residue $Z$,
\begin{align}\label{TF_2_7}
&\delta \Sigma^{R(\ref{fig:2LoopSigma2}a\&b)}(\bk,\Omega_m) = 
\frac{i \omega_m + \Sigma_{\parallel \omega_m} - \varepsilon_{k}}{2 (2 k_F a)^2} 
\\\nn&\times
\log^2\left[ \frac{E_F/(k_F a)}{{\rm max}\{|i \omega_m + \Sigma_{\parallel}(\omega_m) - \varepsilon_k|, \omega^\perp_{\rm FL}\} } \right].
\end{align}

\subsubsection{Two-loop vertex correction}

\begin{figure}
\includegraphics[width=0.4\textwidth]{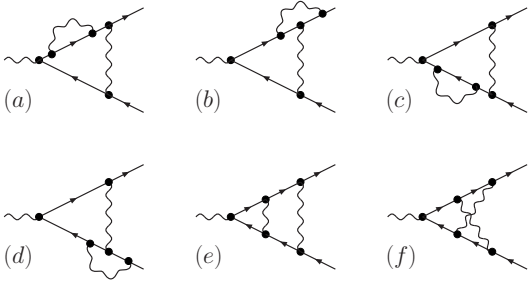}
\caption{Two-loop vertex corrections with exchange of two transverse bosons (wiggly line).}
\label{fig:2LoopVertex}
\end{figure}

We next consider logarithmic singularities in the vertex correction 
at two loop order, arising from the  exchange of transverse bosons.
The corresponding diagrams are presented in Fig.~\ref{fig:2LoopVertex}. 
We find partial cancellations between the contributions, similar to the 
 cancellations between the two-loop diagrams for the self-energy. 
 Using one-loop results for the diagrams in Fig.~\ref{fig:Sigma&Gamma2}, 
we obtained  that  the diagrams  (a) and (c) with self-energy insertions cancel the diagrams
 (b) and (d)  with vertex correction insertions. 
Out of two remaining diagrams, the  important one is the diagram (e) with the 
two ladder-type vertex corrections. For the inner vertex correction we can use the one-loop result (\ref{TF_7}), and the remaining integrals are evaluated in a  manner similar to how Eq. (\ref{TF_2_7}) was obtained.  Performing the calculation, we obtain 
\begin{align}\label{TF_2_8}
\lefteqn{\delta \Gamma^{R(\ref{fig:2LoopVertex}e)}_{\bk,\omega_m}(\bq,\Omega_m) =  \frac{1}{2(2 k_F a)^2} \times}
\\\nn&
\log^2\left[
\frac{E_F/(k_F a)}{
{\rm max}\left\{
| i \omega_m \pm \frac{i\Omega_m}{2} + \Sigma_{\parallel \omega_m \pm \Omega_m/2}- \varepsilon_{\bk\pm \bq/2}|,
\omega_{\rm FL} \right\}}
\right].
\end{align}
The remaining diagram (f) does not have a $\log^2$ term -- it vanishes after the  integration over the direction of bosonic momenta. As a result, Eq.~(\ref{TF_2_8}) is the full result for the vertex correction at two-loop order.

\subsubsection{Polarization corrections}
\label{sec:TransFlucPolCorr}

\begin{figure}
\includegraphics[width=0.4\textwidth]{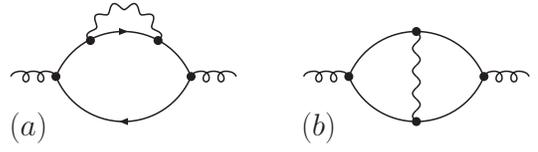}
\caption{Correction to the polarizations due to the exchange of a transverse boson (wiggly line). The external curly line represents either a longitudinal or transverse boson.
}
\label{fig:2LoopTransPol}
\end{figure}

Because the exchange of transverse bosons leads to logarithmic singularities in the electron self-energy, the question arises whether it affects similarly strongly electron polarizations. If it does, 
 logarithmical renormalizations of the self-energies and the polarizations have
 to be analyzed self-consistently and in parallel. We show, however, that this is not the case.  In the evaluation of the self-energy we have already witnessed a partial cancellation of logarithmic singularities arising from self-energy insertions and vertex corrections. We demonstrate below that in the polarization bubble, self-energy and vertex corrections due to a transverse boson exchange conspire so that singular logarithmic terms are exactly canceled out.

The polarization bubble with transverse self-energy and vertex corrections 
 is shown in Fig.~\ref{fig:2LoopTransPol}. Expressing each of the two diagrams in terms of fermionic and bosonic propagators and performing calculations in the same way as before we find that 
 each diagram is logarithmically divergent. However, comparing them, 
 we find after simple manipulations that the integrands  are identical and only differ in the overall sign. The issue is then the interplay between the 
combinatorial factors for the two diagrams. The self-energy can be inserted into the upper or the lower fermion propagator, hence the  diagram (a) has  a combinatorial factor of two. 
The vertex correction  diagram (b) does not contain this factor, however, 
  the logarithm in this diagram comes from the region where the
 internal momentum carried by the transverse boson is sufficiently large
 such that one of the two virtual
 particle-hole pairs,  either before or after the boson exchange, is 
 far away from its mass-shell. This implies that the logarithm in the  diagram (b) comes from two distant regions, and this exactly compensates the combinatoric factor of two in the diagram (a).
 As a result, the logarithmic singularities  from vertex and self-energy insertions to the polarization bubble cancel out, and there is no need to redo the calculations of logarithmical self-energy and vertex corrections.

\subsection{Renormalization group equations}

So far, we have found singular logarithmic renormalizations of the electron residue $Z$ and the vertex due to the exchange of transverse bosons
 in one- and two-loop orders. The two-loop renormalization is exactly $1/2$ of the square of the one-loop renormalization. 
  One can attempt to go beyond perturbation theory and sum up the leading logarithmic corrections up  to infinite order.  The interplay between one-loop and two-loop renormalizations is a good indication that the system is renormalizable  such that one can apply a renormalization group (RG) treatment. For this purpose, we introduce a running residue $Z$ for the electrons and a running
 vertex $\Gamma$,  defined at a certain energy scale $D$. 
Applying a standard procedure of integrating out high-energy degrees of freedom,
 one obtains RG flow equations for $Z$ and $\Gamma$:
\begin{align}
\label{TF_3_1}
\frac{\partial (1/Z)}{\partial \log D} &= \frac{1}{2 k_F a} Z \Gamma^2,
\\
\label{TF_3_2}
\frac{\partial \Gamma}{\partial \log D} &= \frac{1}{2 k_F a} Z^2 \Gamma^3.
\end{align}
 The flow equation (\ref{TF_3_1}) follows from the one-loop result 
for the self-energy, Eq.~(\ref{TF_4}), after taking into
 account that a renormalized electron propagator and 
the renormalized two vertices in Fig.~\ref{fig:Sigma&Gamma2}(a)
 yield an additional factor $Z \Gamma^2$.
 Similarly, the RG equation (\ref{TF_3_2}) follows 
from Eq.~(\ref{TF_7}) upon evaluating the diagram
 \ref{fig:Sigma&Gamma2}(b) with renormalized propagators and vertices. 
The flow starts at an energy scale $D_0 \sim E_F/(k_F a)$ and 
 is cut at either the distance to the fermionic mass-shell or at the scale $\omega_{\rm FL}$ of Eq.~(\ref{TF_5}).
The initial conditions for the RG equations are $Z(D_0) = 1$ and $\Gamma(D_0) = 1$.

One can  easily verify that the combination $\Gamma Z$
 is an invariant of the RG flow,
\begin{align}\label{TF_3_3}
\Gamma Z = 1.
\end{align}
This invariance is the manifestation of the 
 cancellations between vertex and self-energy corrections which we 
observed in the explicit computations above. 
In particular, the condition $\Gamma Z =1$ ensures that the polarizations 
$\Pi_\alpha$ remain unchanged by the RG flow. Indeed,  
in the calculations with logarithmical accuracy, one can select a cross-section with the smallest energies and add self-energy renormalizations and independent  renormalizations of the vertices on both sides of the selected 
cross-section (this is how we get an overall factor of 2 for the diagram 
\ref{fig:2LoopTransPol}(b) in the previous subsection). This implies that the dressed diagram for the polarization bubble differs from the bare one by $(\Gamma Z)^2$ and is obviously not renormalized. 

We emphasize that the relation $\Gamma Z = 1$ is valid to logarithmic accuracy, i.e., only the logarithmically singular parts of the self-energy and vertex correction cancel. Non-logarithmic self-energy and vertex renormalization terms do not cancel, but are small in our theory  $1/(a k_F)$. Exact cancelations due to Ward identities only arise in the limit of small momenta $q/\omega \to 0$. We are considering, however, the opposite limit of small frequencies, $\omega/q \to 0$,
 when there is no exact cancelation between self-energy and vertex diagrams.
 
Using $\Gamma Z =1$ the flow equation for the residue $Z$ can be simplified to
\begin{align}\label{TF_3_4}
\frac{\partial Z}{\partial \log D} &= - \frac{1}{2 k_F a} Z.
\end{align}
This equation is easily solved and we finally obtain for the electron Green function
\begin{align}\label{TF_3_5}
G(\bk,\omega_m) &= \frac{Z\left[i \omega_m + \Sigma_\parallel(\omega_m) - \varepsilon_k\right]}{i \omega_m + \Sigma_\parallel(\omega_m) - \varepsilon_k}
\end{align}
where the energy-dependent $Z$-factor is given by
\begin{align}\label{TF_3_6}
Z(\epsilon) =
&
\left\{
\begin{array}{cc}
\displaystyle \left(\frac{D_0}{\epsilon}\right)^{-\eta_\perp} 
&
{\rm if}\, |\epsilon| > \omega_{\rm FL}
\\
\displaystyle \left(\frac{D_0}{\omega_{\rm FL}} \right)^{-\eta_\perp}
&{\rm if}\, |\epsilon| < \omega_{\rm FL}.
\end{array}
\right.
\end{align}
with $\eta_\perp = 1/(2k_F a)$ and, $D_0 \sim E_F/(k_F a)$.
At the QCP, 
\begin{align}\label{TF_3_5_a}
G(\bk,\omega_m) &\propto  \frac{D_0^{-\eta_\perp}}{(i \omega_m + \Sigma_\parallel(\omega_m) - \varepsilon_k)^{1-\eta_\perp}}.
\end{align}
The form of the electron Green's function close to a nematic QCP,
 given by Eqs.~(\ref{TF_3_5}), (\ref{TF_3_6}), and (\ref{TF_3_5_a}),
 with the longitudinal self-energy $\Sigma_\parallel$ defined in Eq.~(\ref{LF_3}), is the main result of this work. 
Separating real and imaginary parts of $G$, we obtain  Eqs.~(\ref{IN_1}) and (\ref{IN_2}) presented in the Introduction.

 We emphasize that the Green's function contains the two energy scales 
$\omega^\parallel_{\rm FL} \sim E_F/((k_F a)^4 \lambda^3) \propto \xi^{-3}$
 and $\omega_{\rm FL} \sim E_F/((k_F a)^2 \lambda) \propto  \xi^{-1}$
 The self-energy $\Sigma_\parallel$ evolves at
 $|\omega_m| \sim \omega^\parallel_{\rm FL}$ from a FL form at  $|\omega_m| < \omega^\parallel_{\rm FL}$
to a non-FL form at  $|\omega_m| > \omega^\parallel_{\rm FL}$. 
In addition, the Green's function develops an anomalous dimension $\eta_\perp$ if the distance to the mass-shell exceeds the energy $\omega_{\rm FL}$. For 
$k_F a \gg 1$ which we assumed to hold, this anomalous dimension $\eta_\perp$ is small.

\section{Summary and discussion}

We analyzed the electron Green's function in an isotropic metal 
in two spatial dimensions, close to a nematic QCP at which
 the Fermi sphere spontaneously develops a quadrupolar moment. 
 The hallmark of this quantum phase transition is the presence of two critical bosonic modes representing  two quadrupolar polarizations. These two modes are 
 characterized by different dynamics. Whereas the polarization longitudinal to the quadrupolar momentum tensor is damped by particle-hole pairs in the metal and has the dynamical exponent $z_\parallel = 3$, the transverse mode remains undamped with $z_\perp = 2$.\cite{Oganesyan} 

The self-energy correction due to the longitudinal fluctuations, $\Sigma_\parallel$, has been investigated before.\cite{Oganesyan,MRA,Kee,woelfle,charge_we,sslee,Metlitski10,Senthil10} At one-loop order, its singular part depends only on frequency and acquires a strong non-FL form, $\Sigma_\parallel(\omega_m) \sim \omega_m^{2/3}$, for frequencies larger than $|\omega_m| > \omega_{\rm FL}^\parallel \sim \xi^{-z_\parallel}$, where $\xi$ is the correlation length of the transition. In the present work, we focused on the modifications of the electron Green function due to the exchange of transverse bosons. We  performed a systematic perturbative analysis within an effective Eliashberg-type  theory which operates with electron propagators already dressed by the
one-loop longitudinal self-energy $\Sigma_\parallel$, see Eq.~(\ref{LF_3}). 
 This effective theory has to be treated with extra care as it possesses spurious divergencies.~\cite{woelfle_2}  For example, the transverse polarization at the one-loop order does not 
reproduce  the original $z_\perp = 2$ dynamics of the transverse fluctuations. 
However, these divergencies are compensated by terms in the perturbative expansion which are formally of higher order in the number of loops
and the fully renormalized bosonic propagator preserves $z=2$ dynamics. 
This has been first established in a perturbation analysis in Ref.~\onlinecite{woelfle_2}. We extended that analysis to higher orders and also included into consideration the curvature of the FS.  We found that spurious divergencies indeed cancel out at all orders, and the curvature does not affect the cancellation.  We found similar cancellation of transverse renormalizations 
for the fermionic self-energy at the mass-shell.

Our key result is the discovery that the exchange of a transverse fluctuation 
leads to a singular logarithmic correction to the residue $Z$ 
of the electron Green function
 already at one-loop order. 
 The logarithm is cut by either the distance to the mass-shell
 or by the energy scale $\omega_{\rm FL} \sim \xi^{-1}$, whichever is larger. The apperance of the energy scale inversely proportional to $\xi$ is an 
unexpected result because transverse bosons have $z_\perp = 2$,
 and scaling arguments suggests that a characteristic energy scale set by the 
 transverse mode is $\omega_{\rm FL}^\perp \sim \xi^{-2}$.
This  scale has manifestations in thermodynamics,\cite{woelfle_2} but we found that it is  unimportant for the renormalization of $Z$ arising from transverse boson exchange (see, however, the remark below on the renormalization of $\xi$). 
 The reason is that the logarithmic enhancement of $Z$ 
 comes from the region of the phase space where the transverse 
 bosonic propagator is near its pole, i.e., $\Omega_m \sim v_F q *(aq)$, and the upper limit of the $1/q$ behavior of the integrand for $Z$ is $q \sim 1/a$ in which case $\Omega_m \sim v_F q$.  The lower limit is $\xi^{-1}$, and the 
 interplay between the two limits yields the $z=1$ scaling of  
$\omega_{\rm FL}$.

We also found, at one-loop order, a singular logarithmic correction 
to the vertex $\Gamma$.  We extended the calculations to  two-loop order and found $\log^2$ terms, both for the self-energy and the vertex. We then applied RG strategy and obtained flow equations for the running $Z$ and $\Gamma$. We found that the flow equation satisfies $Z \Gamma =1$ and also verified this result in explicit calculations to two-loop order. The condition $Z\Gamma =1$ implies, in particular, that bosonic polarizations remain uneffected by logarithmical singularities although individual self-energy and vertex corrections to the polarization bubble are logarithmically singular.

The solution of the RG equations yields our main result: an electron Green function at the nematic QCP develops the anomalous dimension $\eta_\perp = 1/(2 k_F a)$, where $k_F$ is the Fermi momentum and $a$ is a quadrupolar scattering length. The fully renormalized Green's function is given by Eq. (\ref{IN_1}). Away from QCP the anomalous dimension persists if the distance to the mass-shell exceeds the energy scale $\omega_{\rm FL} \sim \xi^{-1}$. The anomalous dimension is detectable in, e.g.,  MDC ARPES measurements, see Eq.~(\ref{IN_2}).  

There are additional logarithmically singular corrections in the theory that we neglected in our treatment for simplicity. First, it was recently argued by Metlitski and Sachdev\cite{Metlitski10} 
and  Mross {\it et al.}\cite{Senthil10}
that the longitudinal fluctuations also contribute a logarithmically singular correction to the fermionic propagator
 but only in three-loop order. It is an interesting open question as to how these additional logarithms exactly affect the renormalization group flow of the full theory and, more importantly, whether they modify the $z_\perp = 2$ dynamics of the transverse fluctuations or not.

Second, in the mean-field theory the correlation length is given by $\xi \sim 1/\sqrt{1+g_{c,2}}$, where $g_{c,2}$ is the $n=2$ charge Landau parameter  which approaches $g_{c,2} \to -1$ at the QCP. As shown in Ref.~\onlinecite{woelfle_2}, $\xi$ by itself acquires logarithmic corrections when the self-interaction of nematic fluctuations is taken into account. In our present analysis, we also neglected these latter logarithmic corrections. We expect that the additional RG flow for the correlation length is decoupled from the flow of the fermionic residue, $Z$, so that only the expressions for the crossover energies $\omega_{\rm FL}$ and $\omega_{\rm FL}^\parallel$ are affected. Nevertheless, due to this flow of $\xi$ the frequency dependence of the residue of the Green's function should become sensitive to the so far elusive energy scale $\omega^\perp_{\rm FL} \sim \xi^{-z_\perp}$.

\section{Acknowledgment}
We are thankful to D.L. Maslov for his interest to this work and useful comments. We acknowledge helpful discussions with C. Castellani, E. Fradkin, L. Dell'Anna, M. Metlitski, W. Metzner, V. Oganesyan, A. Rosch, S. Sachdev, 
T. Senthil, and  P. W{\"o}lfle. 
This work was supported by the DFG through SFB 608 and FOR  960 (M.G.) and NSF-DMR-0906953 (A. V. Ch.).

\end{document}